\documentclass[reviewcopy]{elsart}
\usepackage{amssymb}
\usepackage{epsfig}%
\usepackage{graphicx}
\usepackage{dcolumn}
\usepackage{bm}
\usepackage{subfigure}
\begin{document}
\begin{frontmatter}
\title{A fluctuating environment as a source of periodic modulation}
\author[unt]{Simone Bianco\corauthref{simo}}
\ead{sbianco@unt.edu}
\corauth[simo]{Corresponding author}
\author[unt,cnr,pisa]{Paolo Grigolini}
\ead{grigo@unt.edu}
\author[paolo]{Paolo Paradisi}
\ead{paradisi@le.isac.cnr.it}
\address[unt]{Center for Nonlinear Science, University of North Texas,
  P.O. Box 311427,Denton, Texas 76203-1427, USA}
\address[cnr]{Istituto dei Processi Chimico Fisici del CNR, Area della
  Ricerca di Pisa, Via G. Moruzzi, 56124, Pisa, Italy}
\address[pisa]{Dipartimento di Fisica "E.Fermi" - Universit\`a di Pisa, Largo  Pontecorvo, 3 56127 PISA}
\address[paolo]{Istituto di Scienze dell'Atmosfera e del Clima (CNR), Sez. di Lecce, Strada Provinciale Lecce-Monteroni, km 1.2, 73100, Lecce, Italy}
\thanks{PG and SB acknowledge ARO and  Welch for financial support 
through Grants W911NF-5-00059 and Grant B-1577, respectively. PP acknowledges 
financial support from MIUR through grant
``Bando 1105/2002'', Project $\# 245$.}

\date{\today}

\begin{abstract}
We study the intermittent fluorescence of a single molecule, jumping from the
"light on" to the "light off" state, as a Poisson process modulated by a
fluctuating environment. We show that the quasi-periodic and
quasi-deterministic environmental fluctuations  make the distribution of the
times of sojourn in the "light off" state depart from the exponential form,
and that their succession in time mirrors environmental dynamics. As an 
illustration,  we discuss some recent experimental results, where the 
environmental fluctuations depend on enzymatic activity.    
\end{abstract}

\end{frontmatter}
\section{Introduction}\label{introduction}
In the last few years there has been an increasing interest for
single-molecule and, more in general, single-system spectroscopy
\cite{silbey,silbey2}. Usually these systems are characterized by intermittent
fluorescence, namely, by sudden jumps between two states, a  \emph{light on}
and \emph{light off} state.
%
%
The Probability Density Function (PDF) of the times of sojourn or Waiting
Times (WTs) $\tau$ in these two states are denoted by the symbols
$\psi_{on}(\tau)$ and $\psi_{off}(\tau)$. In the case
of Blinking Quantum Dots (BQDs) \cite{blinking1,blinking2,blinking3,blinking4}
there is a general agreement that these  are renewal processes, namely, that 
the jumps between the two states have the effect of resetting to zero 
the system's memory.
The PDFs $\psi_{on}(\tau)$ and $\psi_{off}(\tau)$ 
are found to be inverse power laws, whose indexes are, according to Ref.
\cite{kunokuno}, $\mu_{on} = 1.7$ and $\mu_{off} = 1.75$. 

The authors of Ref. \cite{jcp} afford directions on how to distinguish a
Renewal non Poisson Process (RNPP), which is a homogeneous in time process 
 from the apparently equivalent condition 
generated by a Non-Homogeneous Poisson Process (NHPP), \emph{i.e.}, with a
time dependent rate of event productions $q(t)$. 
Let us recall that a (homogeneous) Poisson process is a particular renewal
process where the WT-PDF is an exponential function whose decay rate coincides
with the (constant) rate of event productions.
The physical origin of RNPPs 
%
%
is still the object of research and debate, whereas that of NHPPs
%
%
%
%
requires however the 
identification of dynamics responsible for the rate time dependence. In this 
Letter we consider the recent work of Ref. \cite{flomembon}, which establishes
the influence of the catalytic activity of a fluctuating enzyme 
on the intermittent fluorescence of the fluorogenic 
substrate.  In this case the function $\psi_{on}(\tau)$ is exponential, and 
the complexity of the system is signaled by the function $\psi_{off}(\tau)$, 
which turns out to be a stretched exponential 
\begin{equation}\label{stretched}
\psi_{off}(\tau) = \psi_0 e^{-\left(\frac{\tau}{t_0}\right)^\alpha},
\end{equation}
with $\alpha = 0.15 $. We argue that this is a NHPP,
namely, an example of the second class of processes. According to the earlier
work \cite{jcp,abgp_pre06,pabbg_aip05} a non-exponential function
 can be expressed as the sum of infinitely many exponential functions \cite{beck,note}. 
In Refs.~\cite{jcp,abgp_pre06,pabbg_aip05} this condition was realized by means of
the stochastic modulation of the event production rate $q(t)$. 
In the present case, 
we adopt a different kind of modulation, which is quasi-periodic, and this 
allows us to establish an evident connection with the popular Stochastic 
Resonance (SR) effect \cite{marchesoni}. SR is a phenomenon produced by making
the  event production rate depend harmonically on time, a condition
generating  a periodic WT reordering: the length of the WTs as a function of
the time order reproduces the rate time dependence 
\cite{osman}.  
In this Letter the quasi-periodic modulation 
 has an internal origin and 
is based on the enzyme's conformation changes. The ultimate theoretical 
justification for the adoption of the periodic  assumption is given by the 
work of Winfree \cite{winfree}, who depicts the fundamental biological 
processes as operating through cyclical processes. Thus, we assume that the 
enzyme's conformation changes are periodic, although non-harmonic, functions 
of time.

\section{Slow Modulation}\label{modulation}

We assume that the time duration of the \emph{on} state is negligible, so that
the system, after jumping from the \emph{off} to the \emph{on} state, jumps
back to the \emph{off} state almost immediately.
Thus,  the \emph{on} state is a kind of flashing snapshot, and the distance
between two consecutive snapshots is described   by $\psi_{off}(\tau)$.  Let
us omit for simplicity's sake the subscript \emph{off} and let us assign to
$\psi(\tau)$ the following form:
\begin{equation}
\label{inverse}
\psi(\tau) = \int_0 ^\infty \Pi(q) \psi_q(\tau) dq = \psi_0 
e^{-\left(\frac{\tau}{t_0}\right)^\alpha},
\end{equation}
being $\psi_q(\tau)=q e^{-q\tau}$
the exponential WT-PDF related to the Poisson process with
event production rate $q$ and $\psi_0=\alpha/(t_0 \Gamma(1/\alpha))$ the
normalization constant. 

The quantity $\Pi(q)$ represents the statistical weight of each
$\psi_q(\tau)$. We imagine $q(t)$ to be a slow function of time, so as to produce
many Poissonian events with the same rate $q$. 
This is called slow modulation condition, and it makes it possible to define
$\Pi(q)$ on the basis of the enzyme's fluctuations.
Each enzyme conformation yields a given $q$ and the slower the enzyme's
conformation change, the larger the statistical weight of this rate. 
It is thus obvious that in the slow modulation limit the statistical weight 
$\Pi(q)$ is inversely proportional to the modulus of $dq/dt$:
%
%
%
\begin{equation}
\label{slowrate}
\left| \frac{dq}{dt} \right| = \frac{1}{T \Pi(q)}.
\end{equation}
The parameter $T$ defines a time scale, driving the slope of q(t),
%
%
which can be properly adjusted to fit the experimental results.
In order to derive a periodic behavior for $q(t)$, we assume the rate $q(t)$
to increase in time from a minimum value $q_0$ to a maximum and very large
value $q_f$, from which it decays slowly to the minimum value again, and
so on, thereby confining the stretched exponential distribution to a limited
time range, which can, however, include several decades, depending on the
choice of the parameters $q_0$, $q_f$ and $T$.
The switches between increasing and decreasing patterns of $q(t)$ correspond
to abrupt slope changes at times $n T$, which make $q(t)$ piecewise
smooth, and exactly periodic as well. However, once a WT is selected we leave 
it to stretch beyond the exact time period $T$, thereby making the exactly 
periodic slope change produce random effects and
quasi-periodicity.
%
%

%
%

According to Eq. (\ref{inverse}), $\psi(\tau)$ is the Laplace
transform of the function $f(q)=q \Pi(q)$. Thus,  from the short-time limit, 
where the stretched exponential reads
\begin{equation}
\qquad  \psi(\tau)  \sim \psi_0 (1 - (\tau/t_0)^\alpha),
\end{equation}
using the Tauberian theorem, we find for $\Pi(q)$, 
 at $q \gg 1$, the following asymptotic expression 
\begin{equation}
\label{useit}
\Pi(q) = C q^{-\alpha - 2}. 
\end{equation}
By plugging Eq. (\ref{useit}) within Eq.(\ref{slowrate}), we obtain the
following equations:
\begin{equation}
\label{slowrate2}
\frac{dq}{dt} =\pm \frac{q^{\alpha + 2}}{T C}.
\end{equation}
The integration of Eq. (\ref{slowrate2}) from $t=0$ to $t=T$ yields two
different results depending on the sign:
\begin{eqnarray}
q_{inc}(t) & = & \frac{q_0}{(1-t/T_{inc})^r}\label{solution_rate2}\\
q_{dec}(t) & = & \frac{q_f}{(1+t/T_{dec})^r},
\label{solution_rate}
\end{eqnarray}
referring to the cases of increasing ($+$ sign) and decreasing ($-$ sign) rate,
respectively.
The times $T_{inc}$ and $T_{dec}$ are defined by
\begin{eqnarray}
T_{inc} & = & \frac{T}{1-(q_0/q_f)^{\alpha+1}},\\
T_{dec} & = & \frac{T}{(q_f/q_0)^{\alpha+1}-1}\\
\end{eqnarray}
and the parameter $r$, determining the coefficient $\alpha$ of the stretched
exponential is given by
\begin{equation}
r  =  \frac{1}{\alpha + 1}.
\end{equation}
These analytical results correspond to the correct derivation of the stretched
exponential under the limiting condition of very slow modulation, namely,
under the condition that $T$ is large enough. 
A periodic behavior of $q(t)$ is defined by applying suitable time-shifts of
the kind $t\rightarrow t+nT$ to the rates given in Eqs.~(\ref{solution_rate2})
and~(\ref{solution_rate}). 
%
%

\section{Numerical results}\label{numerical}

The purpose of this section is to assess by means of artificial sequences how 
large $T$ must be for the WT-PDF $\psi(\tau)$ to fit the stretched exponential
form of Eq. (\ref{stretched}). 
On the other hand, we note that the experimental results of Ref. 
\cite{flomembon} do not yield an exactly periodic time ordering. Thus we plan 
to establish
if it is  possible to find values of $T$ that 
make $\psi(\tau)$ fit the stretched exponential condition without producing a 
too markedly periodic time ordering, which would depart qualitatively from the
experimental results of Ref. \cite{flomembon}. 

The numerical algorithm is based on the jump probability in the time interval
$I=[t,t+dt]$: $p(t)=q(t)dt$, $dt$ being
small enough to satisfy the conditions $p(t) \ll 1$ and $q(t)$ nearly constant
on the interval $I$.
%
%
We used the following values of the
parameters: $\alpha = 0.2$, $q_0 = 10$ and $q_f = 10^3$. Thanks to Eq. 
(\ref{inverse}) the only fitting parameter remains $t_0$. 
\begin{figure}
\centering
\includegraphics[scale=0.5]{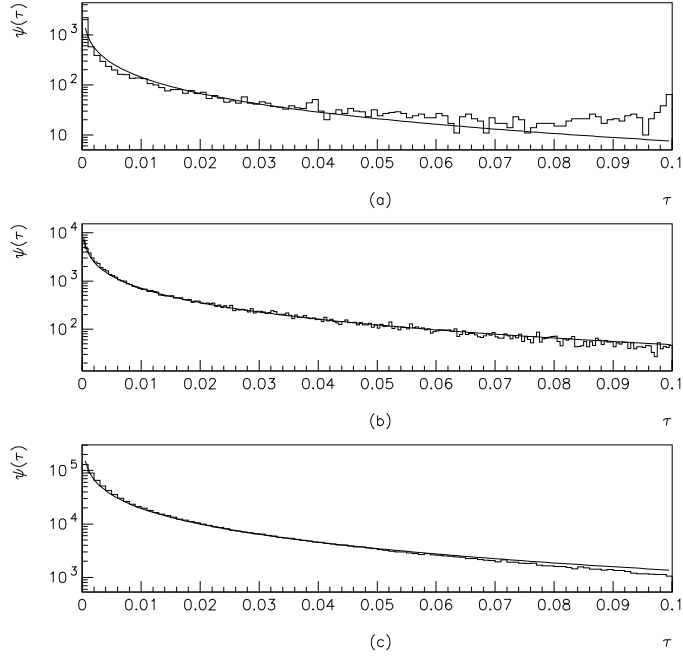}
\caption{The WT-PDF $\psi(\tau)$ as a function of $\tau$. 
(a) $T = 0.1$, (b) $T=1$, and (c) $T=100$. The full line is the stretched
exponential of Eq. (\ref{stretched}), with $\alpha = 0.2$ and $t_{0}$ 
determined by a fitting procedure yielding: (a) $t_0 = 3.1 \cdot 10^{-6}$, 
(b) $t_0 = 4.9 \cdot 10^{-6}$, and (c) $t_0 = 5.1 \cdot 10^{-6}$.}
\label{fig1}
\end{figure}
Figures~\ref{fig1}(a) to~\ref{fig1}(c) show the WT-PDF
$\psi(\tau)$ for different values of the parameter $T$. Fig.~\ref{fig1}(a) 
refers
to $T = 0.1$. In this case the slow modulation condition is violated.
This is the case where the process can be classified as renewal. 

\begin{figure}
  \centering
  \includegraphics[width=9cm,height=5cm]{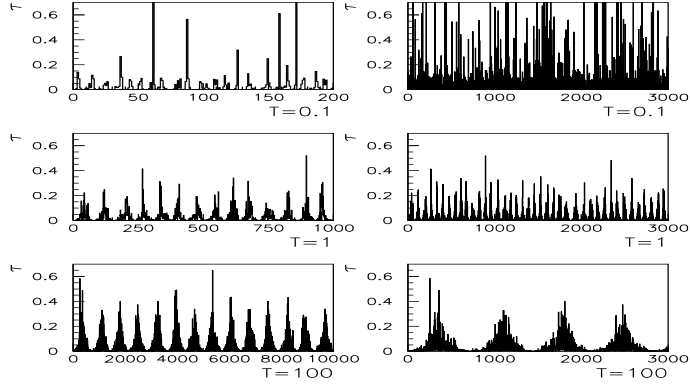}
  \caption{A comparison of the chronological order of WTs for
    different values of the parameter $T$. On the left side we arranged the
    same number of peaks (fixed number of enzyme fluctuations), while on the 
  right side we compare the number of peaks per fixed number of WTs (or 
system's jumps). $T = 0.1, 1, 100$ from the top
   to the bottom. }\label{chr_ord}
  \end{figure}

We see that for both Fig.~\ref{fig1}(b) and Fig. ~\ref{fig1}(c), referring to 
$T=1$ and $T=100$, respectively, the agreement between the numerical result 
and the stretched exponential prediction of Eq. (\ref{stretched}) is accurate
throughout the whole time regime. Both conditions, therefore, are in principle
compatible with the experiment of Ref. \cite{flomembon}. 
To select which of the two conditions is closer to the experiment of Ref.
\cite{flomembon}, let us plot the WTs $\tau_{i}$ according to the chronological
order, i.e., with respect to their order of appearance $i$.
%
%
The results are illustrated in Fig.~\ref{chr_ord}, for different values of the parameter
$T$. We make the comparison among different values of $T$ with two distinct
criteria: in the left column we adopt different numbers of WTs, namely we
refer to different chronological times, so as to produce the same number of
abrupt slope changes of $q(t)$;
 in the right column, on the contrary, the different plots contain the same number of
WTs, namely range from $1$ to the same maximum chronological time.
The results of the right column allow us
to establish a qualitative comparison with the experimental results of Ref. 
\cite{flomembon}. Furthermore, we see that in accordance with the stochastic
 resonance phenomenon \cite{osman}, increasing $T$ has the effect of making 
more evident the resonant character of this effect. We see that the condition 
corresponding to $T = 1$, where the resonant effect is much less marked than 
in the case $T = 100$, produces the best qualitative agreement with the 
experimental result of Ref. \cite{flomembon}. It is worth remarking that in
this condition the WT-PDF $\psi(\tau)$ still retains a markedly stretched
exponential behavior so as to ensure a fully qualitative agreement with the 
experimental results of Ref. \cite{flomembon}.   

\section{Aging Experiment and Conclusion}
\begin{figure}
\centering
\includegraphics[scale=.32,angle=270]{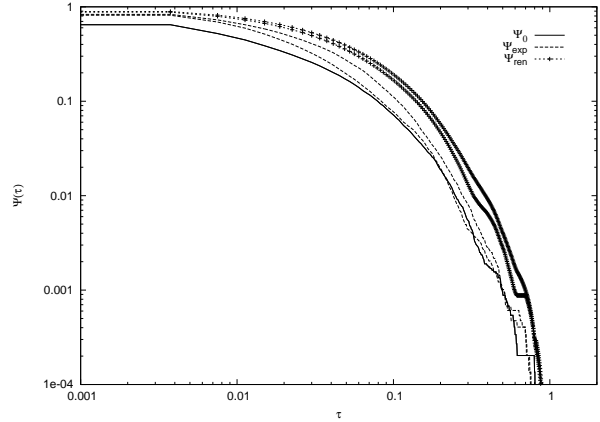}
\caption{Results of the renewal aging analysis on the survival probability,
    for aging times $t_a = 0.1, 6.1$ from the bottom to the top, and 
    parameter $T = 1$. The aging effect is reduced.}
\label{aging1}
\end{figure}

\begin{figure}
\centering
\includegraphics[scale=.32,angle=270]{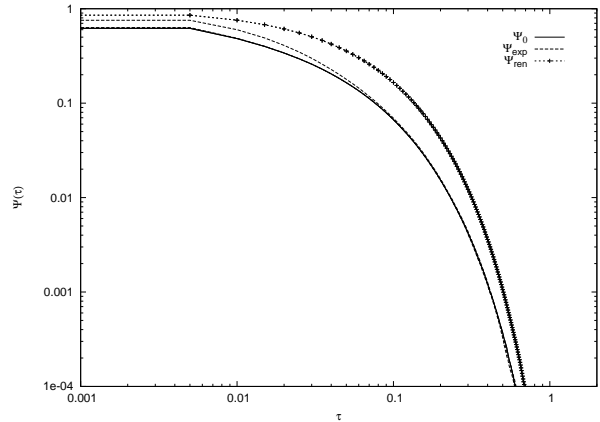}
\caption{Results of the renewal aging analysis on the survival probability,
    for aging times $t_a = 1, 10$ from the bottom to the top, and 
    parameter $T = 100$. The renewal aging is almost absent in this case.}
\label{aging100}
\end{figure}

To draw final conclusions we apply to the artificial sequences $\{t_{i}\}$ of
Section~\ref{numerical} the Aging Experiment (AE) described in the earlier publication of
Ref.~\cite{jcp}. Here we briefly recall that the method is based on  the
computation of the survival probabilities $\Psi_{exp}(\tau,t_{a})$ and
$\Psi_{ren}(\tau,t_{a})$. 
Both survival probabilities are aged and derived from time sequence by delaying 
the beginning of the waiting process for an event until a distance 
$t_a$ from each event of the sequence.  $\Psi_{exp}$ is derived by 
applying this procedure to the time sequence, and $\Psi_{ren}$ is 
derived from the experimental histogram by  assuming that the renewal 
condition holds true.
Upon  increase of $t_{a}$, both $\Psi_{exp}(\tau, t_a)$ and 
$\Psi_{ren}(\tau,t_a)$ can become much slower than $\Psi_{0}(\tau) \equiv 
\Psi(\tau,t_a = 0)$. If $\Psi_{exp}(\tau, t_a) = \Psi_{ren}(\tau, t_a)$, and 
both are slower than $\Psi_{0}(\tau)$, this is the sign that the process 
is renewal and non-Poissonian, insofar as  in the Poisson case, a 
renewal process does not yield aging. In Fig.~\ref{aging1} and Fig.~\ref{aging100}, we show a 
pair of dotted curves, illustrating  $\Psi_{exp}(\tau, t_a)$ at two 
different ages, and a pair of dashed curves, illustrating 
$\Psi_{ren}(\tau, t_a)$ for the corresponding ages.
In both cases the two curves of the latter pair generate a decay 
significantly slower than $\Psi_{0}(\tau)$. This is only a consequence 
of the fact that $\Psi_{0}(\tau)$ is a stretched exponential.  We see, 
however, that in Fig.~\ref{aging1} both dotted curves are slower than 
$\Psi_{0}(\tau)$, although faster than the corresponding dashed curves, 
thereby indicating that a form of aging may exist, albeit 
significantly reduced. This is an indication that many renewal events 
are present~\cite{abgp_pre06}.
In the case of Fig.~\ref{aging100}, on the contrary, the two dotted curves 
virtually coincide with $\Psi_{0}(\tau)$, if we neglect the short-time 
region, thereby implying that the deviation from exponential 
relaxation is caused by a so slow modulation as to cancel all the 
renewal events that may be non-Poissonian. We note that the condition of
reduced aging of Fig.~\ref{aging1} corresponds to $T=1$, which, as shown by
Fig.~\ref{chr_ord}, generates for the chronological order a structure
qualitatively similar to that of the experimental results 
of Ref.~\cite{flomembon}. \vspace{.3cm}\\
\indent We thus conclude that the sequence of WTs
emerging from the experimental results of Ref.~\cite{flomembon} may be a NHPP, generated by
a stochastic resonance produced by the enzyme's conformation changes. The
debate on whether  non-exponential relaxation is due to inhomogeneities or
cooperative molecular motion is of wide interest in biophysics (see, for
instance Ref.~\cite{bishop}).  We note that the recent Letter of Ref.~\cite{flomembino} confirms the
stretched nature of the WT-PDF of Ref.~\cite{flomembon} while demanding further theoretical
progress to understand the correlation among events. Then we notice that
experimental evidence for a periodic behavior of intermittent fluorescence can
also be found in~\cite{baldini}. We think 
that the theory developed in this Letter, along the lines of the 
earlier work of~\cite{osman}, may turn out to be useful to relate the 
experimental results to the enzyme dynamics.



\begin{thebibliography}{100}
\bibitem{note}Beck
discussed energy rather than time distributions and, more important, he focused
on the range of small rates, corresponding to the limit of large energies 
(or times). Here, on the contrary, we focus on the short-time behavior.
\end{thebibliography}
\end{document}